\documentclass[twocolumn]{revtex4}

\usepackage[]{graphicx}

\begin{document}

\title{Lateral distribution of radio signal measured in showers with energy 5$\cdot$ 10$^{16}$-10$^{18}$ eV at the Yakutsk EAS array}


\author{S. Knurenko}
\author{I. Petrov}

\affiliation{ Yu. G. Shafer Institute of Cosmophysical Research and Aeronomy Siberian Branch of the RAoS, Russia.}

\email{s.p.knurenko@ikfia.sbras.ru}

\begin{abstract}
We present of measurements of radio emission from EAS at the frequency 32 MHz performed in 2008-2012. Showers were selected by zenith and azimuth directions, sampled by energy. A mean lateral distribution function was fitted to the data each interval. The paper presents unified approximation formula for all three energy intervals of lateral distribution, using connection between equation (1) and (2) coefficients with energy and geomagnetic angle. Here we analyze the connection between the shape of lateral distribution and the depth of the maximum shower development X$_{max}$ too.  
\end{abstract}


\maketitle

\section{Introduction}

One of the techniques to register ultra-high energy extensive air showers (EAS) is measuring strength of radio pulse by antennas. Unlike traditional techniques, including optic measurements of air shower propagation radio technique can operate in any atmospheric condition except during thunderstorm conditions for whole observation period, which dramatically increases effective time of air showers registration. It is easier to use and much cheaper than other ground detectors in existing air showers array.

The Yakutsk array measured three components of air shower: the total charged component, the muon component and Cherenkov radiation. From these components using average lateral distribution function (LDF) the integral characteristics of air shower, the total number of charged particles, the total number of muons and full flux of Cherenkov light at the sea level are recovered. All these shower characteristics are used for further model-free air shower energy estimation as shown in [1]. Cherenkov light registered at the sea level moreover is used to recover air shower longitudinal distribution and it characteristics, cascade curve and depth of maximum X$_{max}$ [2, 3]. Using this, in future is possible to find a relation between the characteristics of the radio emission and characteristics of the EAS, including slope of the radio emission LDF with depth of maximum, as shown in [4].

\section{Radio Event Selection for Analysis}

For the season 2008-2012 were recorded 600 air shower events with radio emission. Showers energy were above 3$\cdot$10$^{16}$ eV, and zenith angle
$\theta$ $\leq$ 70$^\circ$. For further analysis were selected only 421 showers, appropriate selection criteria of this paper. Therefore, for analysis at the Yakutsk array we use following criteria:
\begin{enumerate}
  \item	The shower selected if ADC prehistory contains radio pulse with amplitude 5 times more than noise level and pulse is localized within time gate equal to delay of “master” from small or large arrays.
  \item Extensive air shower axis must be within perimeter of central array with radius 600 m. Zenith angle $\theta$ $\leq$ 35$^\circ$. Azimuth angle $\phi$ chosen such a way as to exclude influence of polarization effect. That is, the amplitude of the crossed antennas were equal or weren't go beyond limit (3-5) $\%$.
  \item Amplitude is calculated by formula [5]:
    \begin{eqnarray*}
        \varepsilon =
        \pm\sqrt{\left|\frac{1}{N_{Pairs}}
        \sum\limits_{i=1}^{N-1}
        \sum\limits_{j>i}^{N}
        s_{i}[t]s_{j}[t]\right|}
    \end{eqnarray*}
\end{enumerate}

These criteria were used to derive EAS radio emission lateral distribution function (LDF).

With selected events, we plotted dependence of maximum amplitude of radio pulse from distance between air shower axis and antenna (Fig. $\ref{icrc2013-0054-01}$) for three intervals of energy. Energy was determined by Cherenkov detectors data [1]. Analytical expression of experimental data is given by function:

 \begin{figure}[t]
  \centering
  \includegraphics[width=0.4\textwidth]{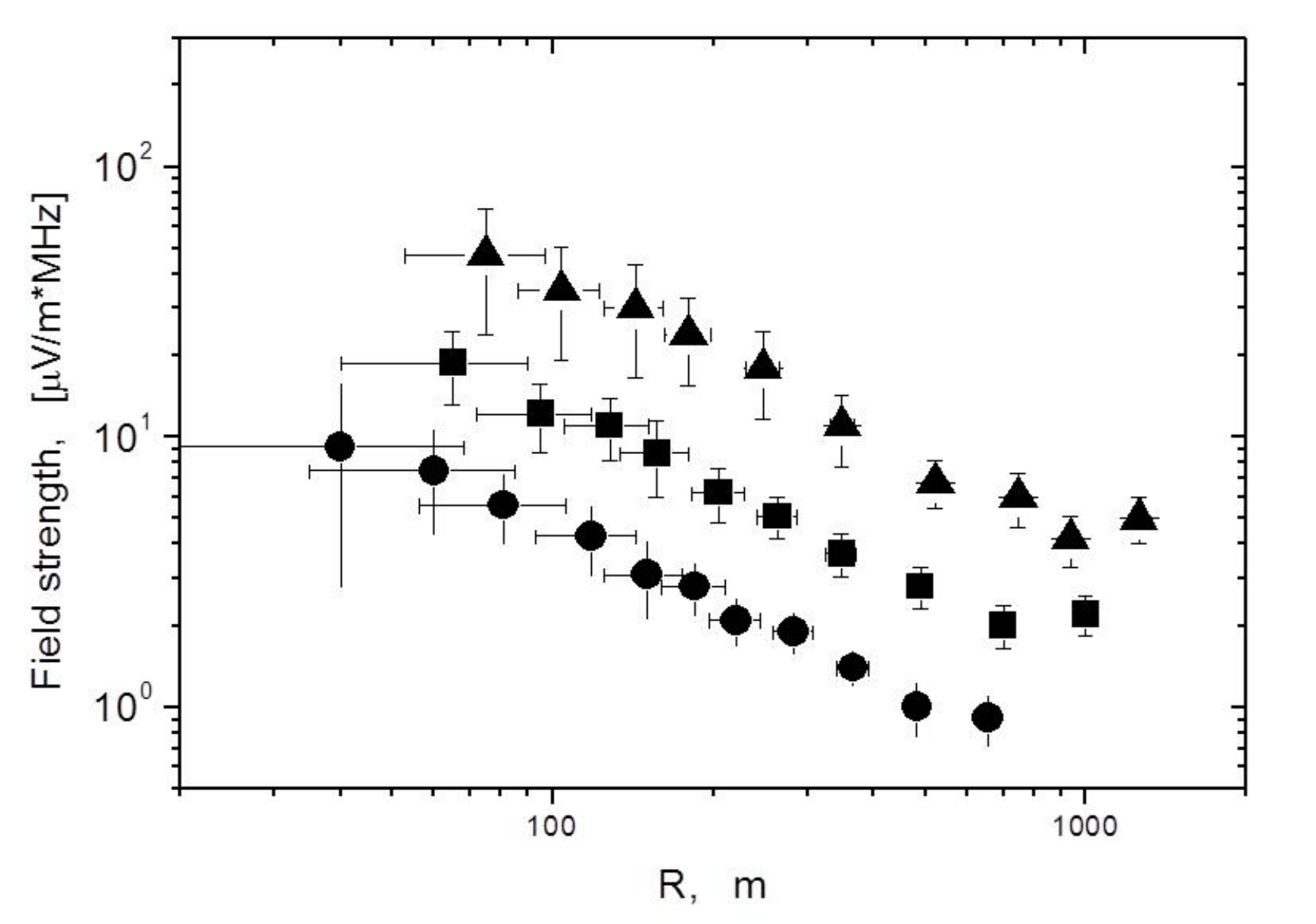}
  \caption{ Averaged lateral distribution of radio emission
   at frequency 32 MHz 1.73$\cdot$10$^{17}$ eV,                4.38$\cdot$10$^{17}$ eV и 1.32$\cdot$10$^{18}$ eV
}
  \label{icrc2013-0054-01}
 \end{figure}

\begin{eqnarray*}
   \varepsilon = k(r/R_{0})^{-b}
\end{eqnarray*}

Each curve was fitted separately. Then we found the dependence of the coefficients from energy and zenith angle (in the case of the Yakutsk array zenith angle is equal to geomagnetic angle)

\begin{eqnarray}
    K(E) = (1.3\pm0.3)(E_{0}/10^{17} eV)^{(0.99\pm0.04)} \\
    b(E,\theta)=(0.81\pm0.25)(1-\cos\theta)^{1.16\pm0.05}
\end{eqnarray}

Fig. $\ref{icrc2013-0054-01}$ shows that slope of average LDF changes with the distance. At large distances from shower axis, radio emission attenuates slowly. From equations (1) and (2) was derived the formula that describes EAS radio emission LDF, which includes dependence form from energy and zenith angle:

\begin{eqnarray}
  \varepsilon(E,\theta,R) = (15\pm1)(1-\cos\theta)^{1.16\pm0.05} \nonumber  \\ \cdot
  \exp\left(-\frac{R}{350\pm25.41}\right)
  \cdot\left(\frac{E_{p}}{10^{17}}\right)^{0.99\pm0.04}
\end{eqnarray}
where, $\theta$ – zenith angle, R – distance to antenna, E$_{p}$ – primary particle energy.

\begin{figure}[t]
  \centering
  \includegraphics[width=0.4\textwidth]{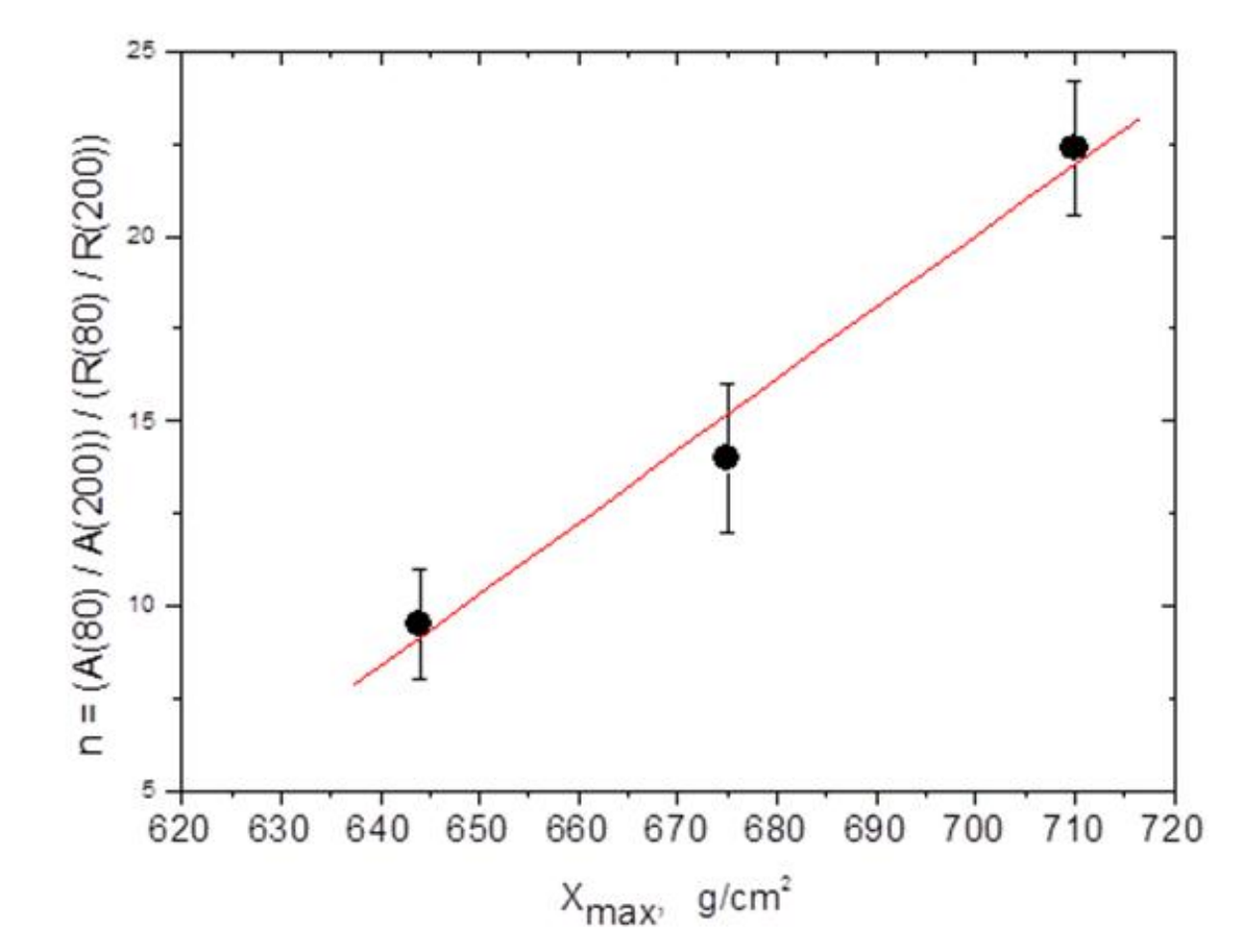}
  \caption{ Dependence of LDF slope n from depth of shower maximum
}
  \label{icrc2013-0054-02}
 \end{figure}

By analogy with [4] and using showers, where the radio emission was measured at the same time, we carried out a comparison of the slope of LDF with depth of maximum of shower X$_{max}$. Fig. $\ref{icrc2013-0054-02}$ shows the slope of radio emission LDF from Xmax. Depth of maximum determined from Cherenkov detectors measurements of the Yakutsk array and slopes determined from ratio of amplitudes at distances 80 and 200 m for three energies: 1.73$\cdot$10$^{17}$ eV, 4.38$\cdot$10$^{17}$ eV and 1.32$\cdot$10$^{18}$ eV.

Fig. $\ref{icrc2013-0054-02}$ implies that slope of air shower radio LDF correlates well with X$_{max}$, which in future can be used for analysis of longitudinal distribution of individual showers and subsequently for an independent estimation of cosmic rays mass composition.

\section{Conclusion}

Measurements of the Yakutsk array showed: a) there is correlation between measured maximum amplitude of radio signal with air shower energy obtained from measurements of main components of EAS at observation level. This follows from joint consideration of radio signal value and air shower energy; b) the form of lateral distribution radio signal depends on depth of air shower maximum X$_{max}$ (Fig. $\ref{icrc2013-0054-02}$).
	
In future, we plan to increase numbers of antennas and use experimental data for determination of X$_{max}$ in individual showers and mass compositions by the method developed in [6].

{\bf Acknowledgements:} We express our thanks to the Ministry of Education and Science of the Russian Federation for the financial support of this study as part of the RFBR 12-02-31442 mol$\_$a.

\end{document}